\newcommand{\g}{$\gamma$}
\newcommand{\cd}{C$_6$D$_6$}

\documentclass[onecolumn]{webofc}
\usepackage{epsfig}
\usepackage{graphicx, animate}
\DeclareGraphicsExtensions{.eps,.png,.jpg,.gif}
\usepackage[varg]{txfonts}   % Web of Conferences font
\usepackage{booktabs}
\usepackage{array} %% needed for advanced table manipulation
\newcolumntype{L}[1]{>{\raggedright\let\newline\\\arraybackslash\hspace{0pt}}m{#1}}
\newcolumntype{C}[1]{>{\centering\let\newline\\\arraybackslash\hspace{0pt}}m{#1}}
\newcolumntype{R}[1]{>{\raggedleft\let\newline\\\arraybackslash\hspace{0pt}}m{#1}}

\graphicspath{{graphics/}{graphics/arch/}{Graphics/}{./}} % Look in these folders for graphics
\begin{document}
\title{New detection systems for an enhanced sensitivity in key stellar (n,$\gamma$) measurements}
\author{
J.~Lerendegui-Marco\inst{1}\footnote{\email{jorge.lerendegui@ific.uv.es}} \and %
V.~Babiano-Su\'arez\inst{1} \and %
J.~Balibrea-Correa\inst{1} \and %
C.~Domingo-Pardo\inst{1} \and %
I.~Ladarescu\inst{1} \and %
A.~Tarife\~no-Saldivia\inst{1} \and %
V.~Alcayne\inst{2} \and %
D.~Cano-Ott\inst{2} \and %
E.~Gonz\'{a}lez-Romero\inst{2} \and %
T.~Mart\'{\i}nez\inst{2}
% ...
and the n\_TOF Collaboration}
\institute{%
Instituto de F\'{\i}sica Corpuscular, CSIC - Universidad de Valencia, Spain \and
Centro de Investigaciones Energ\'{e}ticas Medioambientales y Tecnol\'{o}gicas (CIEMAT), Spain
% ...
}

\abstract{%
Neutron capture cross-section measurements are fundamental in the study of astrophysical phenomena, such as the slow neutron capture (s-) process of nucleosynthesis operating in red-giant and massive stars. However, neutron capture measurements via the time-of-flight (TOF) technique on key $s$-process nuclei are often challenging. 
Difficulties arise from the limited mass ($\sim$mg) available and the high sample-related background in the case of the unstable $s$-process branching points. Measurements on neutron magic nuclei, that act as $s$-process bottlenecks, are affected by low (n,$\gamma$) cross sections and a dominant neutron scattering background. Overcoming these experimental challenges requires the combination of facilities with high instantaneous flux, such as n\_TOF-EAR2, with detection systems with an enhanced detection sensitivity and high counting rate capabilities. 
This contribution reviews some of the latest detector developments in detection systems for (n,$\gamma$) measurements at n\_TOF, such as i-TED, an innovative detection system which exploits the Compton imaging technique to reduce the dominant neutron scattering background and s-TED, a highly segmented total energy detector intended for high flux facilities. The discussion will be illustrated with results of the first measurement of key the $s$-process branching-point reaction $^{79}$Se(n,$\gamma$).}
\maketitle

\section{(n,$\gamma$) measurements for the $s$-process: challenges and solutions}\label{sec:challenges_solutions}
Neutron capture reactions play a fundamental role in the slow neutron capture (s-) process of nucleosynthesis operating in red-giant and massive stars~\cite{Kappeler11}, which is responsible for the formation of about half of the elements heavier than iron. However, neutron capture measurements via the time-of-flight (TOF) technique on key $s$-process nuclei are challenging. As a consequence, for many nuclei, uncertainties in the stellar (n,$\gamma$) cross sections are still significantly larger than the very precise abundance observations~\cite{Kappeler11}.

Among the relevant isotopes that still present large uncertainties, one can identify neutron-magic nuclei which act as $s$-process bottlenecks~\cite{Domingo:22}. These nuclei feature very low (n,$\gamma$) cross sections and the accuracy of the TOF experiments is limited by the dominant neutron scattering background. Sec.~\ref{sec:iTED} describes a new development aimed at suppressing this background and enhancing the detection sensitivity.

Experimental efforts are also focused on the measurement of unstable nuclei which act as branchings of the $s$-process and yield a local isotopic pattern which is very sensitive to the physical conditions of the stellar environment~\cite{Kappeler11}. The limited sample mass available and the high background induced by the sample activity represent the major challenges to experimentally access the (n,$\gamma$) cross sections of these isotopes~\cite{Domingo:22}. To overcome these limitations, new facilities with higher neutron fluxes are needed. Indeed, this was the aim of building n\_TOF-EAR2~\cite{Weiss15}, that thanks to its flight path of only 20 m, became a world-leading facility in terms of instantaneous flux. Moreover, after the recent upgrade of the spallation target, an additional increase of 30-50\% in the flux is expected~\cite{Lerendegui:22_ND}. In order to profit from the high flux of EAR2, one requires also a new generation of radiation detectors with higher granularity, as discussed in Sec.~\ref{sec:sTED}.

\begin{figure}[!b]
  \centering
  \includegraphics[width=0.495\columnwidth]{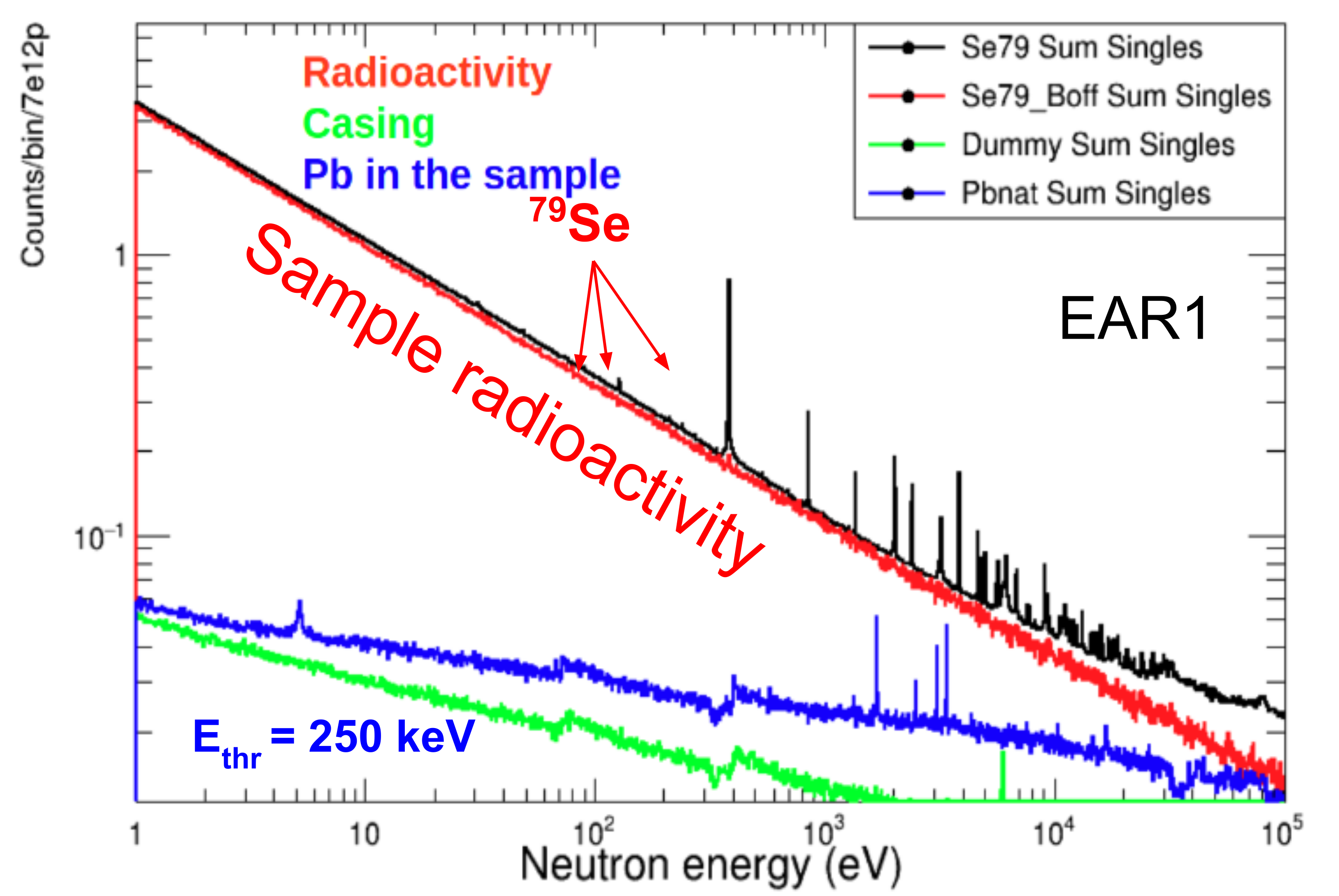}
  \includegraphics[width=0.495\columnwidth]{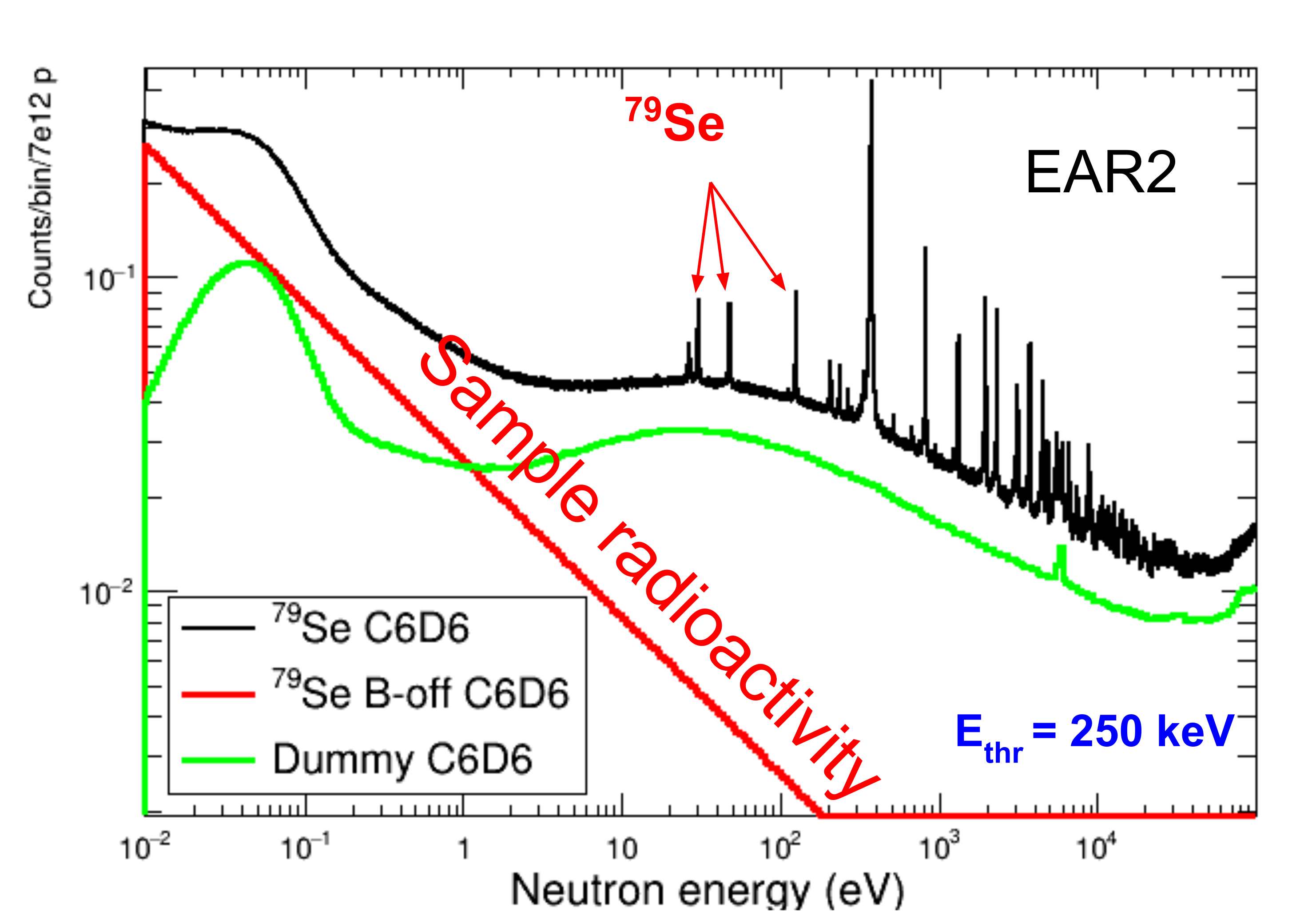}
  \caption{Total counts and background components as a function of the neutron energy measured with the PbSe~($^{78}+^{79}$Se) sample in EAR1 (left) and EAR2 (right). The first resonances of $^{79}$Se are highlighted.}
 \label{fig:Se79HighFlux}
\end{figure}
Recently, the first TOF measurement on the unstable $^{79}$Se has been proposed and carried out at CERN n\_TOF~\cite{Lerendegui21}. The branching at $^{79}$Se is particularly well suited for determining the thermal conditions of the stellar environment thanks to the strong thermal dependence of its beta decay rate~\cite{Kappeler11}. For this experiment, 2.7~mg of $^{79}$Se were produced by means of neutron irradiation of an enriched $^{208}$Pb$^{78}$Se alloy-sample in the high-flux reactor at ILL~\cite{Chiera22,Lerendegui21}. The relevance of having a higher instantaneous flux has been clearly observed in this measurement (see Fig.~\ref{fig:Se79HighFlux}). While in EAR1~\cite{Guerrero13}, the total count rate is dominated by the sample radioactivity, in EAR2~\cite{Weiss15}, thanks to the enhanced (n,$\gamma$)-to-activity ratio in EAR2, $^{79}$Se resonances are clearly visible. Some preliminary results of this measurement are discussed in Sec.~\ref{sec:Se79}. 

\section{i-TED: suppressing n-induced background via \g-ray imaging}\label{sec:iTED}
Neutron-capture time-of-flight (TOF) measurements at CERN n\_TOF have been usually carried out with detection systems based on liquid scintillators, such as C$_{6}$D$_{6}$, which are particularly convenient because of their fast time-response and low intrinsic sensitivity to scattered neutrons~\cite{Plag03}. However, they present limited background rejection capabilities. In particular, in TOF capture experiments on nuclei with low capture-to-scattering ratio (see Sec.~\ref{sec:challenges_solutions}) a large background component arises from scattered neutrons that get subsequently captured in the surroundings of the \cd~detectors~\cite{Zugec14}. This background has represented the dominant contribution in previous (n,\g) experiments in the energy range of interest for nucleosynthesis studies~\cite{Tagliente13}.

\begin{figure}[!htb]
  \centering
  \includegraphics[width=0.33\columnwidth]{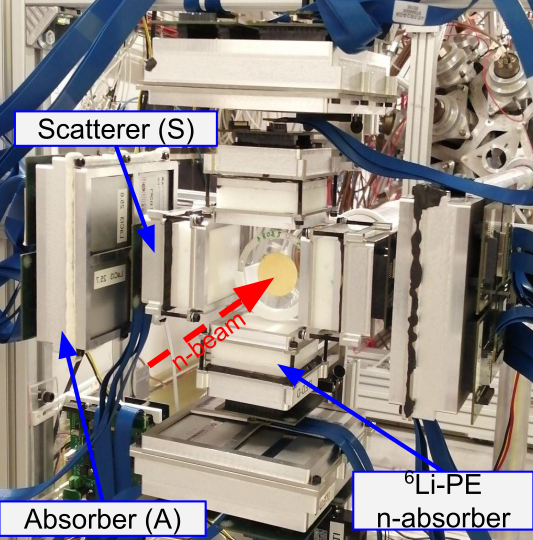}
\includegraphics[width=0.48\columnwidth]{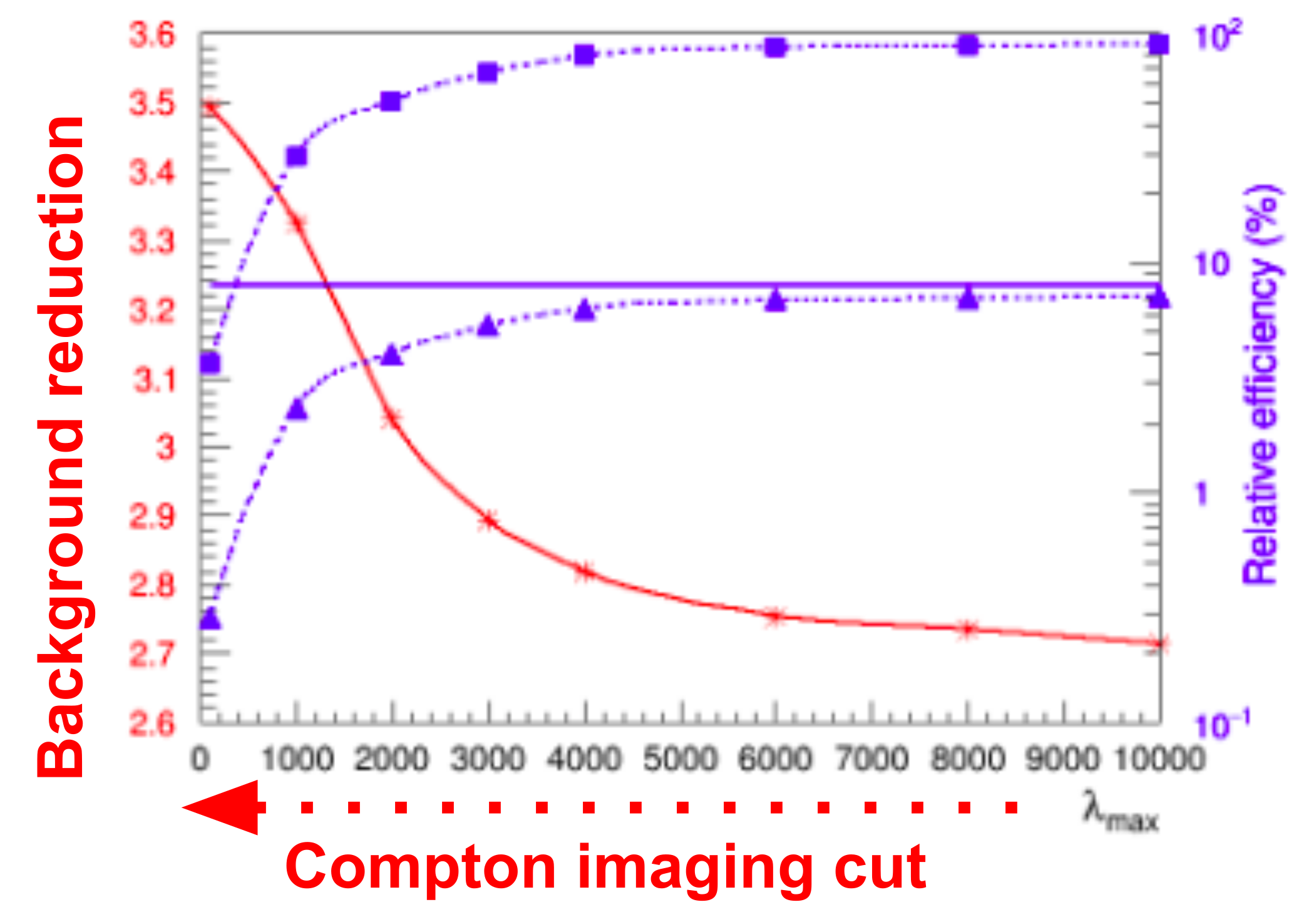}
  \caption{Left: Final i-TED array used in the recent $^{79}$Se($n,\gamma$) experiment at CERN n\_TOF EAR1. Right: Background reduction factor at 10~keV and relative (n,$\gamma$) efficiency as a function of the Compton imaging cut (see text for details).}
 \label{fig:iTED}
\end{figure}

    To reduce this dominant source of background, a system based on Total-Energy Detection (TED) with $\gamma$-ray imaging capability, so-called i-TED, has been recently proposed~\cite{Domingo16}. i-TED exploits the Compton imaging technique with the aim of determining the direction of the incoming $\gamma$-rays. This allows the rejection of events not originating in the sample, thereby enhancing the signal-to-background ration (SBR). This novel detection system has been fully developed and optimized in the recent years~\cite{Babiano20,Balibrea:21}. The final i-TED array (see Fig.~\ref{fig:iTED}) consists of 4 Compton cameras comprising in total 20 LaCl$_{3}$ crystals, and it has been used in 2022 at n\_TOF EAR1 for the aforementioned $^{79}$Se($n,\gamma$) reaction measurement.

The feasibility of the proposed background rejection method was experimentally demonstrated with an early i-TED prototype by measuring the $^{56}$Fe(n,$\gamma$) reaction at CERN n\_TOF-EAR1~\cite{Babiano21}. A background reduction factor of 3.5 was achieved with respect to state-of-the-art \cd~detectors (see red line in the right panel of Figure~\ref{fig:iTED}). The major drawback of the imaging selection is the drop of relative efficiency after applying an imaging cut, shown with blue lines. The line with squares corresponds to the efficiency relative to the situation with no imaging cuts while the line with triangles shows the efficiency in coincidence mode relative to the one of the scatter crystal operated in single mode, which has a maximum value of 8\% indicated with the solid blue line. New methods based on Machine Learning (ML) algorithms have been applied to overcome this limitation in efficiency. The reader is referred to Ref.~\cite{Babiano21} for the details.

\section{s-TED: segmented detection volumes for high flux facilities}\label{sec:sTED}

The high instantaneous flux of n\_TOF-EAR2, which has been further enhanced in the latest upgrade of the facility~\cite{Lerendegui:22_ND}, induces counting rates beyond 10~MHz in the existing \cd~detectors (0.6-1 L volume)~\cite{Plag03}. Moreover, the intense \g-flash arriving to the experimental hall, leads to severe experimental difficulties~\cite{AlcayneND:22}. Among others, counting-rate and time-of-flight dependent variations of the photo-multiplier (PMT) gain and large pile-up effects have been observed.
\begin{figure}[!htb]
  \centering
  \includegraphics[width=0.45\columnwidth]{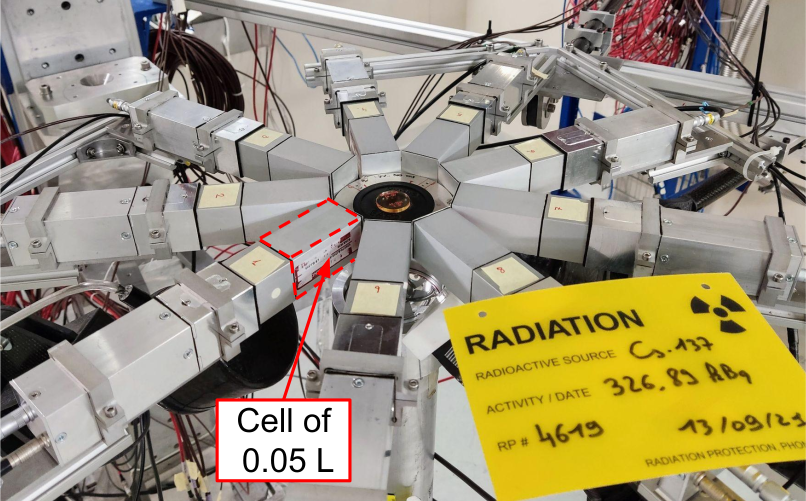}
  \includegraphics[width=0.542\columnwidth]{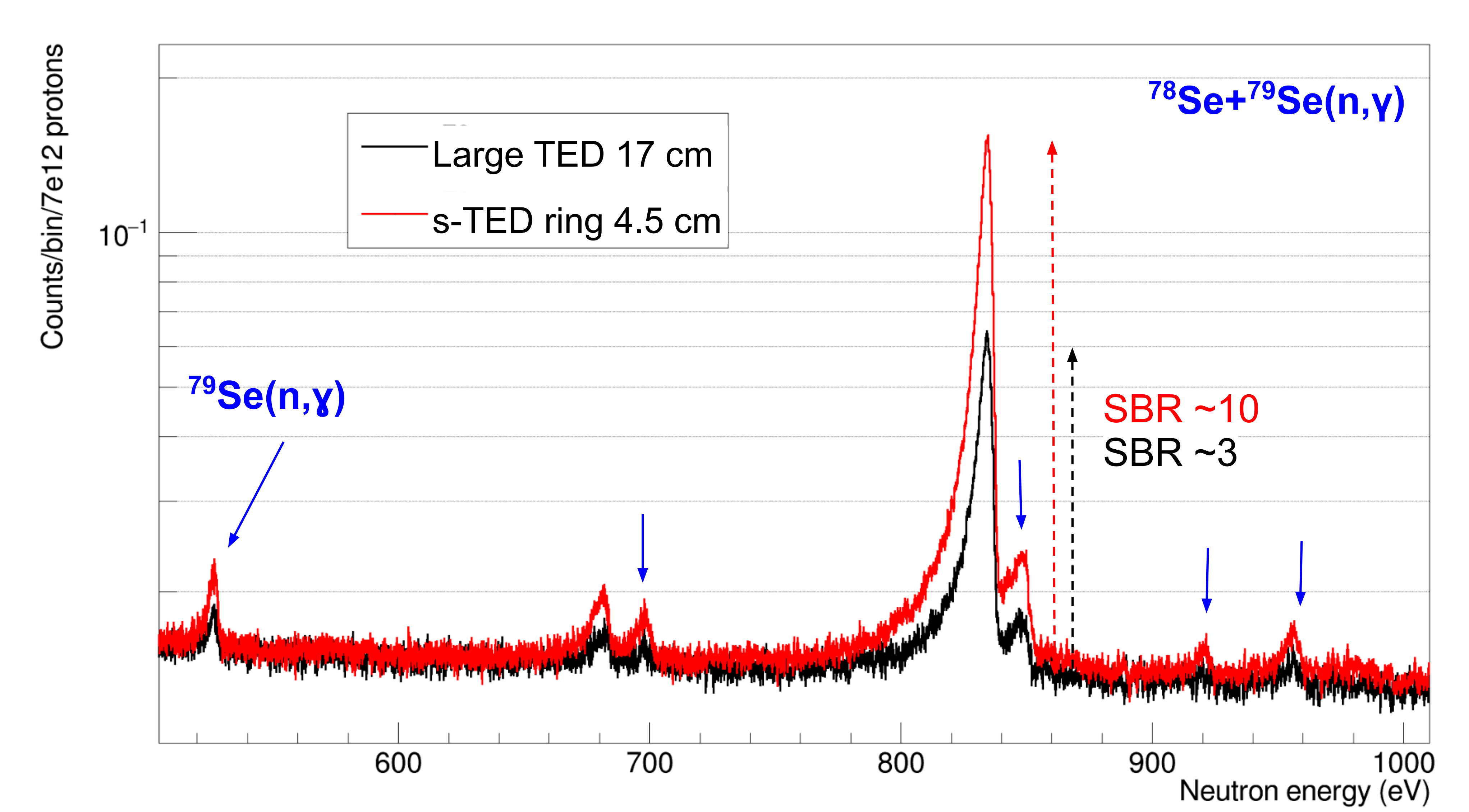}

  \caption{Left: s-TED detector cells in the innovative ring configuration used in EAR2 (right). Right:(n,$\gamma$) counting rate measured at EAR2 with the PbSe($^{78+79}$Se) sample in beam using 2 large \cd~detectors (black) and 9 s-TED cells in ring configuration (red).}
  \label{fig:sTED}
\end{figure}

In order to overcome the aforementioned limitations of conventional TEDs based on relatively large \cd~volumes, a segmented array of small-volume \cd~detectors, so-called s-TED, has been developed~\cite{AlcayneND:22_b}. Each detection cell contains only 49~ml scintillation liquid, 12-to-20 times less than previous designs. The original idea of s-TED was to replace each of the conventional \cd~detectors by an array of 3$\times$3 cells~\cite{AlcayneND:22_b} in order to preserve a similar overall efficiency. To achieve an optimum efficiency and SBR for the (n,$\gamma$) measurements on the unstable $^{94}$Nb~\cite{Balibrea_NPA-X:22} and $^{79}$Se, the s-TED cells have been arranged in a compact-ring configuration around the capture sample shown in the left panel of Fig.~\ref{fig:sTED}. This innovative setup minimizes the distance (4.5~cm) to the capture sample under study, and thus enhances the sensitivity in a significant manner compared to the larger conventional \cd~detectors placed at 17 cm~\cite{Balibrea_NPA-X:22}, as it is shown in the right panel of Fig.~\ref{fig:sTED}. The larger SBR achieved with the ring of s-TED cells has been key to identify the weak $^{79}$Se(n,\g) resonances measured in a sample containing only 2.7~mg of $^{79}$Se embedded in 3.9~g of a eutectic lead-selenide ($^{208}$Pb$^{78}$Se) sample.

\section{First results of $^{79}$Se(n,$\gamma$) using new detection systems}\label{sec:Se79}
The first (n,$\gamma$) measurement on $^{79}$Se has been just carried out at CERN n\_TOF using the two new detection systems described in this work. Choosing the best combination of detection system and experimental area is a key aspect for the success of challenging capture measurements on unstable targets. 

Given the small amount of $^{79}$Se and the high activity of the sample, n\_TOF-EAR2 was the best solution to achieve good statistics and minimize the sample activity background (see Sec.~\ref{sec:challenges_solutions}). On the other hand, 99.7\% of the Se in the sample was $^{78}$Se, for which a high resolution measurement was carried out at n\_TOF EAR1~\cite{Lederer:17}. The latter facility was thus better suited for a reliable assessment of the $^{78}$Se contribution. Besides the radioactivity, a high background was expected from neutron scattering due to the large $^{208}$Pb content (3~g) in the sample. To suppress both sources of background, the usage of the innovative i-TED array was the best solution. This system applies imaging to reduce the n-induced background, as described in Sec.~\ref{sec:iTED}, and thanks to the LaCl$_{3}$ crystals it features a very good energy resolution (~5\% FWHM at 662~keV), which enables to set precise energy selections to reduce the activity background. However, the maximum acquisition rate of its data acquisition system is of 500 kEvents/s which represents, at this time, a limitation for its use at EAR2. For all of the above, i-TED was chosen for the measurement at EAR1, while the measurement at EAR2 was carried out with the new s-TED detectors to fully exploit the high flux of the facility and optimize the sensitivity, as discussed in Sec.~\ref{sec:sTED}.

\begin{figure}[!htb]
  \centering
  \includegraphics[width=0.47\columnwidth]{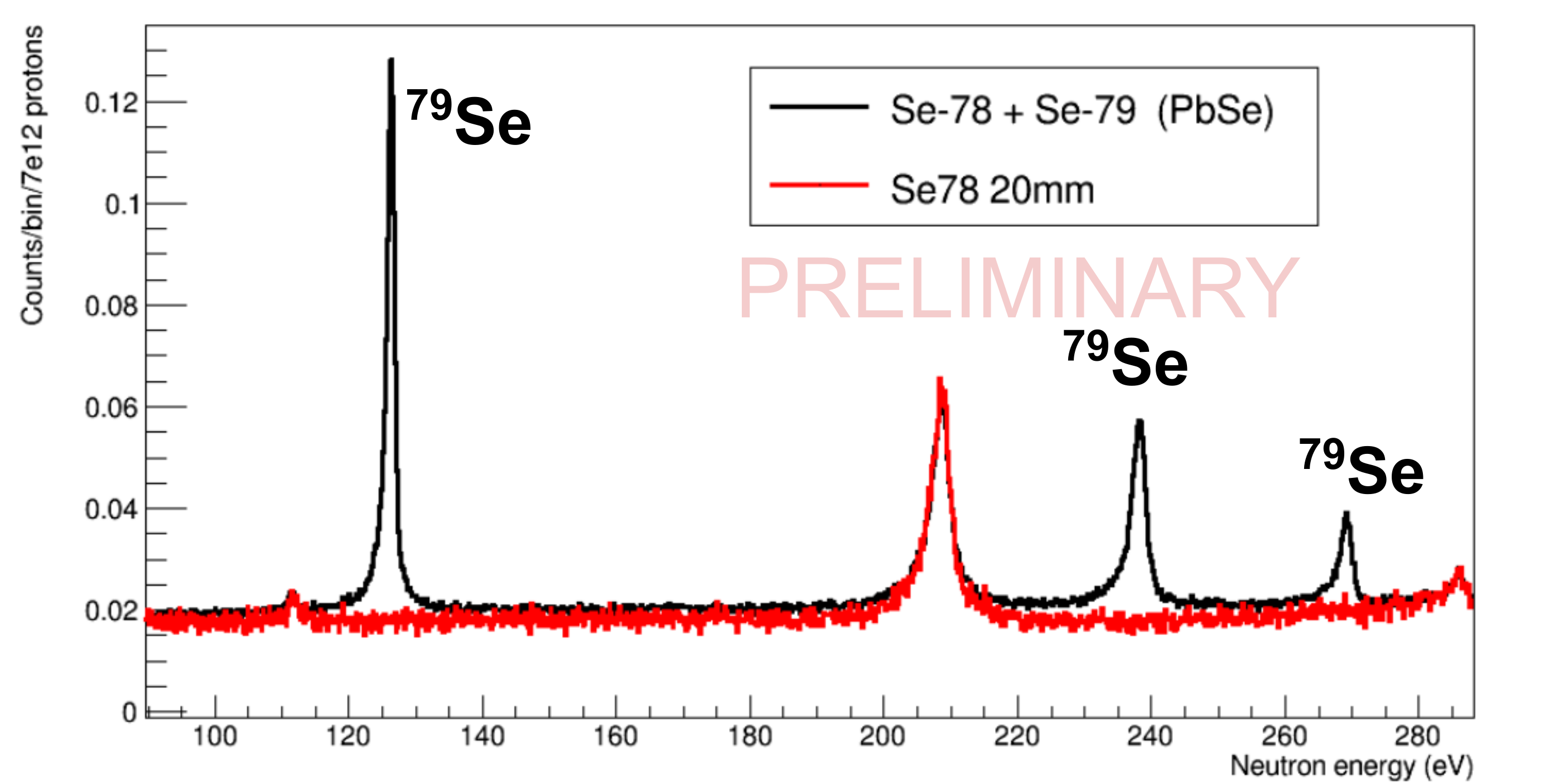}
  \includegraphics[width=0.52\columnwidth]{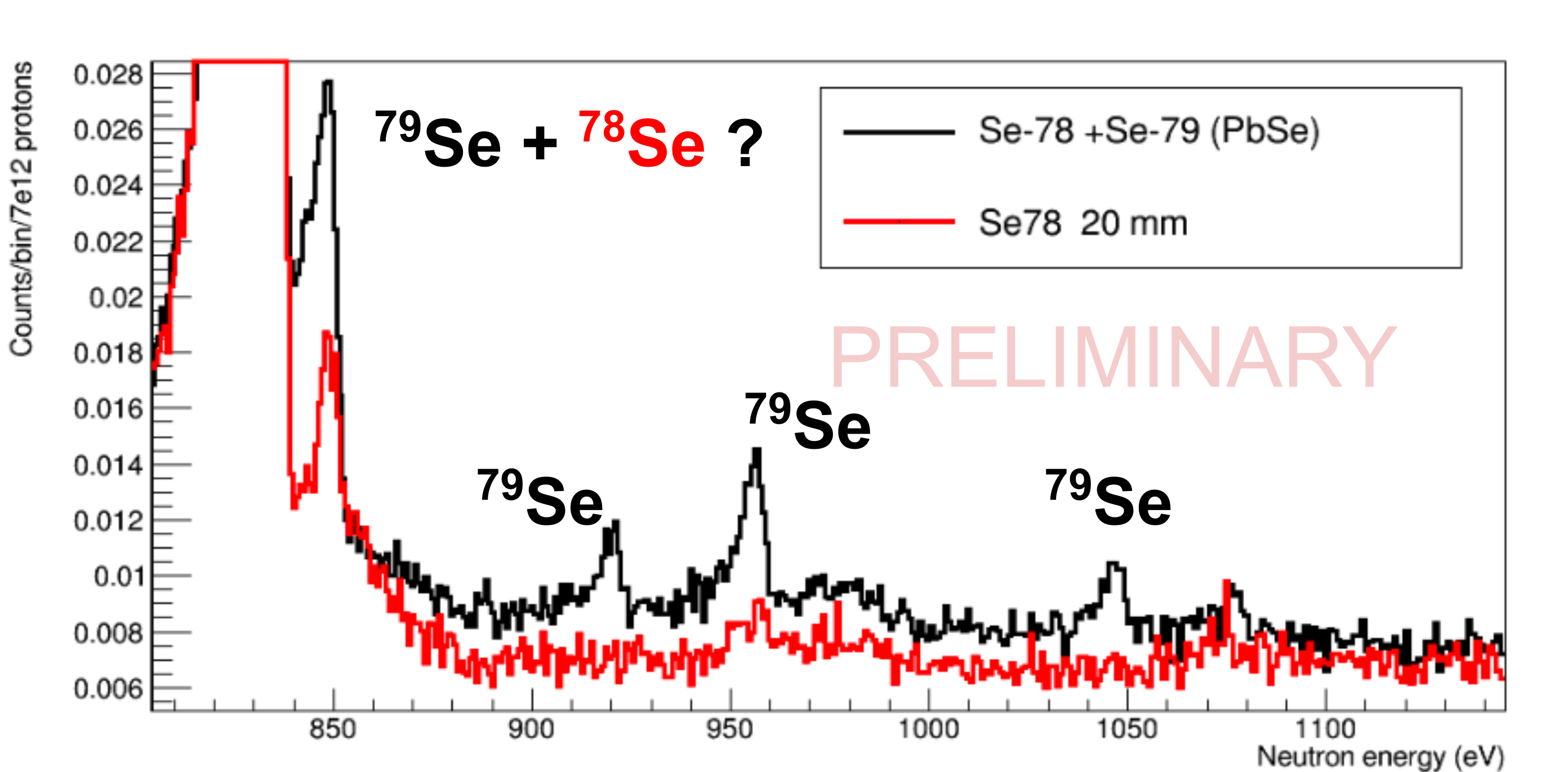}
  \caption{Preliminary (n,$\gamma$) counting rates measured at EAR2 with the PbSe($^{78+79}$Se) sample in two neutron energy ranges.}
  \label{fig:ResonancesSe79}
\end{figure}

 The preliminary analysis and background subtraction of the measurement carried out at EAR2 show promising results. The obtained (n,$\gamma$) counting rate on the PbSe($^{78+79}$Se) sample compared to an ancillary measurement of a pure $^{78}$Se sample (see Fig.~\ref{fig:ResonancesSe79}) indicates that between 10 and 15 capture resonances in $^{79}$Se+n have been measured for the first time in the energy range from 30~eV to 2.5 keV. 

\section{Summary and outlook}\label{sec:summary}
This work has presented recent detector R\&D developments carried out at CERN n\_TOF, that try to solve existing limitation and shortcomings in previous approaches used for determining (n,\g) cross sections of relevance for the $s$-process. i-TED applies Compton imaging aimed at improving the signal-to-background ratio for measurements affected by large neutron-induced backgrounds, such as the neutron magic nuclei acting as $s$-process bottlenecks. New facilities with higher instantaneous neutron flux, such as n\_TOF-EAR2, facilitiate the TOF capture measurements on radioactive $s$-process branching nuclei. However, exploiting the full potential of this facility requires new detectors, such as s-TED, a new array of very small-volume C$_6$D$_6$ detectors, capable of dealing with the counting rate conditions and optimizing the signal-to-background ratio. 

These two novel detection systems have been used in the challenging (n,\g) measurement on the key $s$-process branching $^{79}$Se. The R-Matrix analysis of the final yield, followed by a statistical analysis will allow the calculation of the semi-empirical cross section up to 300~keV, from which the MACS at different $k_{B}T$ can be determined. This will provide the first experimental constraint to actual spread of theoretical calculations of the $^{79}$Se MACS compiled in KaDoNiS. 

\section*{Acknowledgements}
This work has been carried out in the framework of a project funded by the European Research Council (ERC) under the European Union's Horizon 2020 research and innovation programme (ERC Consolidator Grant project HYMNS, with grant agreement No.~681740). J.~Lerendegui-Marco and J.~Balibrea were supported, respectively, by grants FJC2020-044688-I and ICJ220-045122-I funded by MCIN/AEI/ 10.13039/501100011033 and by European Union NextGenerationEU/PRTR. The authors acknowledge support from the Spanish Ministerio de Ciencia e Innovaci\'on under grants PID2019-104714GB-C21, FPA2017-83946-C2-1-P, FIS2015-71688-ERC, CSIC for funding PIE-201750I26. In line with the principles that apply to scientific publishing and the CERN policy in matters of scientific publications, the n\_TOF Collaboration recognises the work of V. Furman and Y. Kopatch (JINR, Russia), who have contributed to the experiment used to obtain the results described in this paper. 

%\bibliography{references}

\end{document}